\documentclass[14pt]{article}
\pdfoutput=1
\usepackage{amsmath,amssymb,amsfonts,amsthm,bm,bbm,cancel,wasysym}
\usepackage{epsfig,graphics,graphicx,epstopdf}
\usepackage{array,booktabs,colortbl,colordvi,multirow}
\usepackage{colordvi,color,xcolor}
\usepackage{hyperref}
\usepackage{rotating}

\usepackage{verbatim}
\usepackage{cite}
\usepackage{subfig}
\usepackage{setspace}
\usepackage{url}
\usepackage[percent]{overpic}
\usepackage{slashed}
\usepackage{authblk}
\usepackage{xspace}
\usepackage{fullpage}
\usepackage{hyperref}

\def\ie{{\it i.e.}}
\def\eg{{\it e.g.}}
\def\etc{{\it etc}}

\def\to{\rightarrow}

\newskip\zatskip \zatskip=0pt plus0pt minus0pt
\def\matth{\mathsurround=0pt}
\def\lsim{\mathrel{\mathpalette\atversim<}}
\def\gsim{\mathrel{\mathpalette\atversim>}}
\def\atversim#1#2{\lower0.7ex\vbox{\baselineskip\zatskip\lineskip\zatskip
  \lineskiplimit 0pt\ialign{$\matth#1\hfil##\hfil$\crcr#2\crcr\sim\crcr}}}

\begin{document}


\begin{flushright}
SLAC-PUB-260601\\
\today
\end{flushright}
\vspace*{5mm}

\renewcommand{\thefootnote}{\fnsymbol{footnote}}
\setcounter{footnote}{1}

\begin{center}

{\Large {\bf Portal Matter and Scotogenic-like Dirac Neutrino Masses}}\\

\vspace*{0.75cm}

{\bf Thomas G. Rizzo}~\footnote{rizzo@slac.stanford.edu}

\vspace{0.5cm}

{SLAC National Accelerator Laboratory}\\ 
{2575 Sand Hill Rd., Menlo Park, CA, 94025 USA}

\end{center}
\vspace{.5cm}


\begin{abstract}
\noindent  
Loops of portal matter (PM) fields, carrying both Standard Model (SM) and dark charges, can generate the necessary kinetic mixing (KM) between the ordinary and dark photons (DP) in vector portal 
scenarios thus allowing for interactions between visible and dark sector fields.  Here we show that the field content of a previously considered model based on a partial $E_6$-like UV-completion of 
such setups can generate light Dirac neutrino masses in the interesting range, $\sim 0.05$ eV, at the one-loop level similar to what happens in scotogenic dark matter (DM) scenarios. While 
general $E_6-$like PM scenarios have been shown to be easily probed at colliders, such as the HL-LHC, uniquely testing this specific subclass of these setups in a direct fashion is found to 
be somewhat more challenging.
\end{abstract}

\vspace{0.5cm}
\renewcommand{\thefootnote}{\arabic{footnote}}
\setcounter{footnote}{0}
\thispagestyle{empty}
\vfill
\newpage
\setcounter{page}{1}



\section{Introduction and PM Model Background}

Understanding the nature of Dark Matter (DM) and its interactions with the particles of the Standard Model (SM) poses one of our current greatest challenges. The literature is full of models 
which attempt to explain what DM {\it might} be and how it may interact with the SM to obtain the observed value of the relic density\cite{Planck:2018vyg}. Of these many scenarios, a 
particularly attractive possibility is the kinetic mixing (KM)/vector boson portal scenario\cite{KM,vectorportal,Gherghetta:2019coi} - a setup which posits (in its simplest manifestation) the existence 
of a new $U(1)_D$ gauge group, having a gauge coupling $g_D$,  with the corresponding gauge boson being termed the dark photon (DP),  
$V$\cite{Fabbrichesi:2020wbt,Graham:2021ggy,Barducci:2021egn,Alonso-Gonzalez:2025xqg,Jorge:2026bbs,Caputo:2026pdw}. This DP will couple directly to DM since the DM carries a non-zero 
dark charge, $Q_D\neq 0$, but the SM fields, \ie, the fermions $u_{L,R},d_{L,R}, e_{L,R}, \nu_L$ as well as the Higgs and the $W^\pm,Z,\gamma$ gauge bosons, at tree level will {\it not} couple to 
$V$ as they are all neutral under $U(1)_D$, \ie, they have $Q_D=0$. Here as part of our setup to be discussed below it will 
be assumed that the DP obtains a non-zero mass in the familiar manner via the spontaneous breaking of this $U(1)_D$ by one or more dark Higgs fields which may in some cases also transform 
non-trivially under the SM gauge symmetries\cite{Li:2024wqj}. In such setups, it is possible that other particles (fermions and/or bosons) carrying {\it both} the SM as well as $U(1)_D$ quantum 
numbers might exist and we have referred to such fields generically as portal matter (PM) in our earlier work\cite{Rizzo:2018vlb,Rueter:2019wdf,Kim:2019oyh,Rueter:2020qhf,Wojcik:2020wgm,Rizzo:2021lob,Rizzo:2022qan,Wojcik:2022rtk,Rizzo:2022jti,Rizzo:2022lpm,Bauer:2022nwt,Wojcik:2022woa,Carvunis:2022yur,Verma:2022nyd,Rizzo:2023qbj,Wojcik:2023ggt,Rizzo:2023kvy,Rizzo:2023djp,Rizzo:2024bhn,Ardu:2024bxg,Rizzo:2024kzu,Rizzo:2025tap,Tewary:2025vij,Rizzo:2025fsy,Rizzo:2026dsj}. 
Such fields will quite naturally be present if the $U(1)_D$ were to be embedded into some larger non-abelian gauge group structure whose breaking at the few TeV scale or above will also 
simultaneously generate the masses for the PM. 
In the case of fermionic PM, these fields are required to be vector-like (VL) with respect to the SM gauge group, $G_{SM}=SU(3)_c\times SU(2)_L \times U(1)_Y=3_c2_L1_Y$, in order to avoid a 
number of unitarity, precision electroweak and Higgs width/production cross section  
constraints\cite{CarcamoHernandez:2023wzf,CMS:2024bni,Alves:2023ufm,Banerjee:2024zvg,Guedes:2021oqx,Adhikary:2024esf,Benbrik:2024fku,Albergaria:2024pji,Chen:2017hak,
Biekotter:2016kgi}. The PM fields are seen to play a vital role in these KM setups: depending upon the details of their SM transformation properties, loops of PM 
particles can induce a sizable KM between the 
DP and the neutral SM $W_3$ and/or $B$ gauge fields via 1-loop vacuum polarization-like graphs. These loops will manifest at low energies, far below the weak scale, as the desired KM of the SM 
photon and $V$.  Further, when the DP and DM are both in the $\lsim 1$ GeV mass range (as we will assume here) the strength of this PM loop-induced KM, denoted by the parameter $\epsilon$, 
can be {\it calculated} to be in the required range to allow 
the DM to reach its observed relic abundance via the familiar thermal freeze-out processes, as in the case of WIMPS\cite{Arcadi:2017kky,Roszkowski:2017nbc,Arcadi:2024ukq}, while at the same 
time satisfying other numerous experimental constraints.  This happens because the charged fields of the SM will pick up a small coupling to the DP as a result of this KM with a 
strength $e\epsilon Q_{em}$, 
with $e$ being the usual charge of the proton.  In standard versions of this setup, no {\it other} additional interactions between the DP (and hence DM) with the SM are expected. However, it has been 
shown that PM loops in models with enlarged dark sector gauge groups, \eg, $G_D=SU(2)_I \times U(1)_{Y_I}=2_I1_{Y_I}$ which breaks at roughly the $\sim 10$ TeV scale, in analogy to 
the SM and as considered in earlier work\cite{Rueter:2019wdf}, can lead to such 
new interactions in the form of dark moments. Note that such enlarged dark gauge sectors necessarily will also imply the existence of enlarged dark Higgs sectors to generate the relevant masses 
and break the symmetries as required. 
Given an enlarged dark sector gauge group such as $G_D$ above, it is quite natural to consider a more complete structure in the UV at least partially combining both $G_D$ as well as 
$G_{SM}$ which {\it ab initio} contains most if not all of the necessary pieces. The analysis below is based on a slightly modified version of one such scenario\cite{Rueter:2019wdf}, inspired by the 
group $E_6$\cite{Hewett:1988xc}, which we have employed in many earlier works in one form or another. We note that the full UV version of this model is incomplete and so this setup is essentially 
a toy example of what might be possible in such completions.

In addition to the properties of DM, the nature (\ie, Dirac vs. Majorana), size and origin of the neutrino masses also remain a significant mystery. Here we will consider the possibility that neutrinos 
are Dirac fields thus predicting the absence of any potential signals/constraints arising from neutrinoless double-beta decay\cite{Shickele:2026fbg}. A basic pair of questions we'd have to address 
in such a scenario are ($i$) if neutrinos are `ordinary' Dirac states then why aren't their masses 
larger, similar to those of the charged leptons, and ($ii$) how do such apparently small Dirac masses get generated within the above $E_6$-like PM model context?  As we will see below, these 
issues are addressed through the existence of both new gauge (\ie, $G_D$) and discrete symmetries which modify the expected properties of $\nu_R$ and which forbid the generation of the Dirac 
mass term through the SM Higgs. $\nu_R$ will not only be odd under this new $Z_2$ discrete symmetry but will also carry a non-zero\cite{nonz} but 
unspecified value of the dark charge $Q_D(\nu_R)=q$.
Instead, this Dirac mass is found to be generated at 1-loop via the exchange of dark sector Higgs scalars as well as neutral fermions which carry dark sector quantum 
numbers and which can be odd under a discrete $Z_2$ symmetry in a manner somewhat (but not entirely) analogous to what occurs in scotogenic models\cite{scoto1} that can also lead to small Dirac 
neutrino masses\cite{scoto2} in the desired range 
$\sim 0.05$ eV. In addition to loop factors, the Dirac mass is then found to be suppressed by the fact that the PM fermion fields are rather heavy, $\gsim 1$ TeV, and that only the vevs, 
$\lsim 1 $ GeV,  which are responsible for $U(1)_D$ breaking and generate the the DP mass will appear therein. Unfortunately, as we will also see below, outside of the generic $E_6-$like 
components of this model, the specific aspects of this setup which are of 
most direct importance for this Dirac mass generation will be shown to be the most difficult to test directly, \eg, via new particle production at colliders. These include not only observation of the new 
non-SM interactions for $\nu_R$ 
(which, while remaining a SM singlet, is not a singlet under $G_D$) but also the direct production of the new dark sector particles in the loop that are specifically responsible for the Dirac neutrino 
mass generation. This issue is made particularly more difficult in some cases by their lack of any conventional SM interactions.

The outline of this paper is as follows:  After this Introduction, in Section 2 we will present/review the specific details of the $E_6$-inspired PM model introduced earlier as are now applicable to the 
problem of generating a small Dirac neutrino mass at the 1-loop level whose resulting range of values is discussed in Section 3. This is a semi-realistic toy model to demonstrate the feasibility 
of this outcome but does not contain all the elements necessary for a complete UV model. Section 4 contains a discussion of the possible UV tests of this scenario via new particle production 
at colliders where we face the somewhat difficult problem that in the present setup all of the fields responsible for this neutrino mass generation are electrically neutral with most of them being 
complete SM singlets. Finally, a discussion of the results and our conclusions are presented in Section 5.


\section{$E_6$-like Model Basics For Dirac Neutrinos}

\subsection{Review}

As noted above, the fermion content of the $E_6$-like PM model that we will consider below (for simplicity, only for a single generation) will consist of the familiar fields contained in the 
fundamental representation of $E_6$, 
\ie, the ${\bf 27}$, {\it plus} an additional $G_{SM}=3_c2_L1_Y$ and $2_I$ chiral singlet, $S_L$, and whose properties under the various symmetries are as summarized in Table \ref{fermtab}. 
This ${\bf 27}$ consists of the the 15 chiral components of the SM fields, $u_{L,R},d_{L,R},e_{L,R}, \nu_L$, which together form the usual ${\bf {bar 5}}+{\bf 10}$ of $SU(5)$, plus the $SU(5)$ singlet 
RH-neutrino, $\nu_R$; recall that when all of these fields are grouped together they form a ${\bf 16}$ of $SO(10)$.  In addition to these, from the 
SM perspective, the ${\bf 27}$ also contains the 10 degrees of freedom in the vector-like PM (VL) fermion fields, $N_{L,R}, E_{L,R}, D_{L,R}$, which form a ${\bf 5}+{\bar {\bf 5}}$ of $SU(5)$ as well as 
a ${\bf 10}$ of $SO(10)$; an additional $SU(5)$ and $SO(10)$ singlet, $S_R$ is also present. In the model below this field together with $S_L$ will obtain a large mass similar to that of the PM fields. 
We note that this is the minimal, partially-unified PM setup required to achieve the desired goal of small Dirac neutrino masses. Note that in the $SU(6)\times SU(2)$ decomposition of 
$E_6$\cite{Hewett:1988xc}, $\nu_R,S_R$ together form a doublet under the $SU(2)$ and will do so again in the discussion but with their roles interchanged. We further 
note that this choice is not unique and that the opposite assignment of the the 3rd component for the $2_I$ isospin is also possible.  For practical purposes here, we need only directly focus on the 
color-singlet fields appearing in this setup as they alone participate in generating the Dirac neutrino mass. Further, looking at the Table, we see that, unlike all of the other 
SM and PM fermions, the purely dark $2_I$ doublet, $(\nu_R,S_R)$, is also odd also under a new discrete $Z_2$ symmetry\cite{Dev:2026ddq}.  
We emphasize again that the full $G_{SM}\times G_D=3_c2_L1_Y2_I1_{Y_I}$ gauge group structure that we consider below does {\it not} 'fit' into $E_6$ (or even into $E_6 \times U(1)$), but 
does so in the absence of $1_{Y_I}$. Hence,  this explains 
the `$E_6$-like' nomenclature employed here; note that while $1_{Y_I}$ commutes with $G_{SM}$ it does {\it not} commute with the full $E_6$ group. 
It is important to remember throughout that while the $2_I1_{Y_I} \to 1_D$ breaking occurs at the scale of roughly $\sim 10$ TeV, $1_D$ itself, corresponding to the region of the DP and DM masses, 
only breaks at the $\lsim 1$ GeV scale.

Even though this is a toy model example, as is well-known, one of the advantages of using $E_6$ as part of the construction is that the SM plus PM fermion fields will automatically 
form an anomaly free set with 
respect to the SM gauge group; here that anomaly freedom is seen to also extend to $G_{SM}\times 2_I$ even in the presence of new field $S_L$. Given the minimal field content above, however, 
$I_{Y_I}$ remains anomalous. We can (and must!) easily include additional fermion fields here, as was done in our earlier work\cite{Rueter:2019wdf}, 
particularly ones that are SM singlets which only transform 
non-trivially under $G_D$, resolving this difficulty. Note that, none of these fermions that we have introduced can act as DM and so we know such additional fields are required if, \eg, the DM 
is fermionic, but no further ones are necessary for the discussion below. For the present analysis we can remain completely agnostic with respect to the exact nature of DM and the further details 
of the dark sector since the knowledge of their natures is not required for this particular aspect of the setup considered below.

We again emphasize that beyond the addition of $S_L$ and the new Higgs fields that are necessary to generate its mass as well as to allow it to decay in a manner similar to the more familiar PM 
fields, the particle content in the setup employed here to generate light Dirac neutrino masses is essentially just that of the vanilla $E_6-$like model as was studied previously\cite{Rueter:2019wdf}.

\begin{table}[h] 
\begin{center}
\caption{Minimal fermionic field content of the present $E_6$-like setup - additional SM singlet dark sector fermion fields will also exist.}\label{fermtab}
\vspace{-.2cm}
\begin{tabular}{ l  c  c c  c  c  c  c  c}
\hline		
 & & SU(3)$_c$ & $T_{3L}$  & $Y$/2 & $T_{3I}$  & $Y_I$/2 & $Q_D$ & $Z_2$  \\ 
\hline
& $\begin{pmatrix} u \\ d\\\end{pmatrix}_L $  & {\bf 3} & $\begin{pmatrix} 1/2 \\ -1/2 \\ \end{pmatrix}$ & 1/6 & 0 & 0 & 0 & + \vspace{.1cm} \\
 & $u_R$ & $\bf{ 3}$ & 0 & 2/3 & 0 & 0 & 0 & + \vspace{.1cm}\\
 & $e_R$ & $\bf{ 1 }$ & 0 & -1 & 0 & 0 & 0 & +\vspace{.1cm}\\
& $\begin{pmatrix} \nu~~N \\ e~~E\\\end{pmatrix}_L $ & $\bf{ 1 }$ & $\begin{pmatrix} 1/2 \\ -1/2 \\ \end{pmatrix}$ & -1/2 & (1/2,  -1/2) & -1/2 & (0, -1) & +\vspace{.1cm}\\
 & $(D,~d)_R$ & $\bf{ 3 }$ & 0 & -1/3 & (1/2, -1/2) & 1/2 & (1,~0) & +\vspace{.1cm}\\
& $\begin{pmatrix} N \\ E\\\end{pmatrix}_R $ & $\bf{ 1 }$ & $\begin{pmatrix} 1/2 \\ -1/2 \\ \end{pmatrix}$ & -1/2 & 0 & -1 & -1 & + \vspace{.1cm}\\
 & $D_L$ & $\bf{ 3 }$ & 0 & -1/3 & 0 & 1 & 1 & +\vspace{.1cm}\\
 & $S_L$ & $\bf{1}$ &0& 0& 0& q -1&q-1  & +      \vspace{.1cm}\\
 & $(\nu,~S)_R$ &$\bf{1}$ &0& 0 &(1/2,  -1/2)& q-1/2 &(q,~q-1)& - \vspace{.1cm}\\
\hline
\end{tabular}
\end{center}
\end{table}

\subsection{Higgs and Tree-Level Fermion Masses}

In order to break the various gauge symmetries thus generating all of the gauge boson and fermion masses - including those for the SM neutrinos - as well as the appropriate hierarchy of the three  
mass scales, $\sim 10$ TeV, $\sim 100$ GeV and 
$\sim$ 1 GeV, several different Higgs fields are required which will transform non-trivially under $G_{SM}$ and/or $G_D$ as well as the above-mentioned $Z_2$ discrete symmetry. Much of this 
symmetry breaking structure is already determined by the lone requirement that we must generate the Dirac masses of the SM and PM {\it charged} fermions (\ie, $u,d,e,D,E$) at the appropriate mass 
scales as was seen in our previous work on these $E_6$-like setups\cite{Rueter:2019wdf}. As noted in this earlier work, fine-tuning is clearly necessary at this level of the analysis to obtain the desired 
vev hierarchy structure. While the colored fermions will not participate in the Dirac neutrino mass generation process itself and so be of limited concern in the discussion that follows, some of the 
Higgs fields that are needed to generate their masses will, however, be seen to be of some more general relevance.

\begin{figure}[htbp] 
\hspace*{0.1cm}\centerline{\includegraphics[width=5.5in,angle=0]{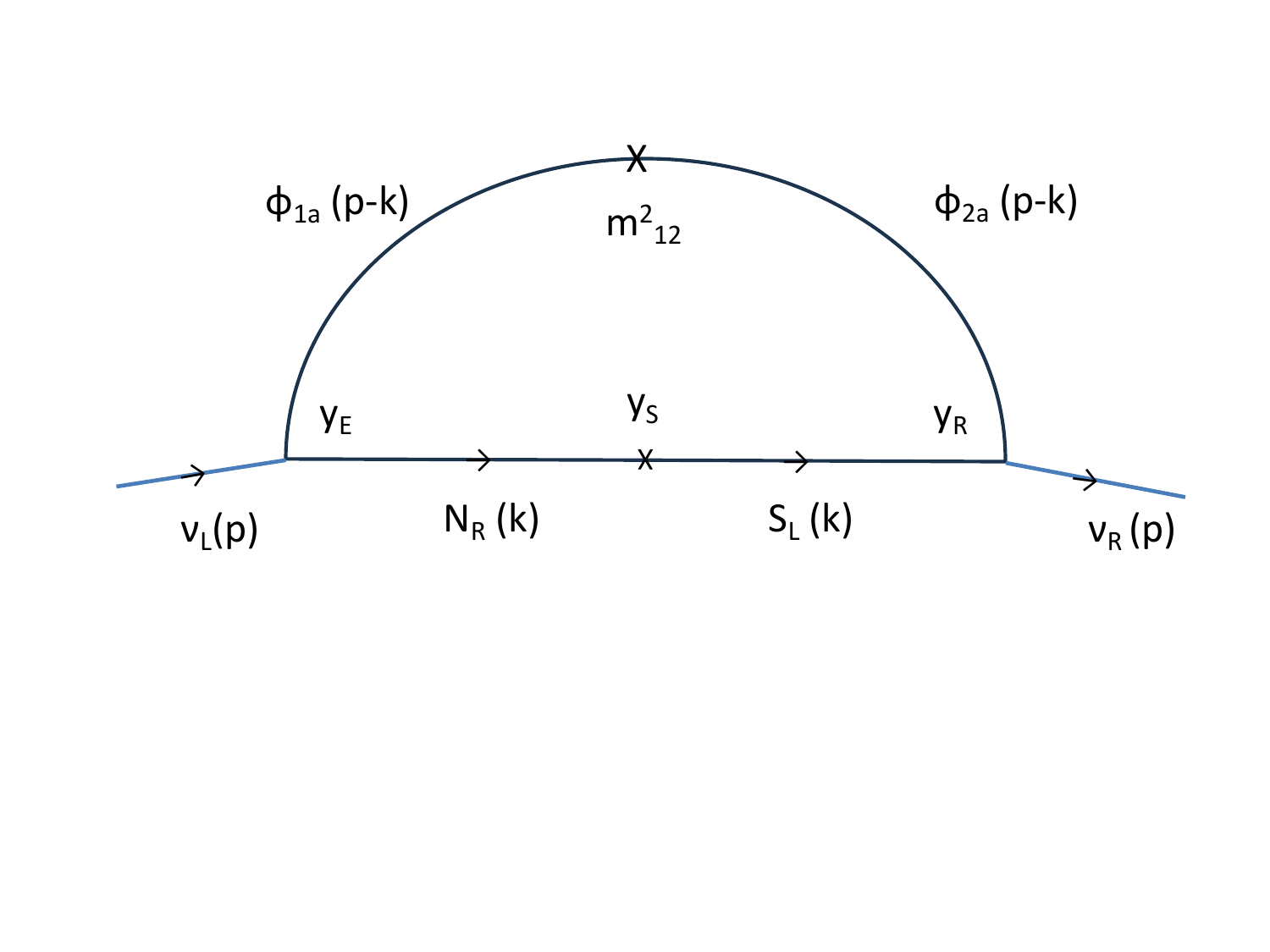}}
\vspace*{-4.0cm}
\caption{Diagram leading Dirac neutrino mass generation in the $E_6$-like PM model discussed in the text at the 1-loop level. The various scalar and fermion fields, Yukawa couplings and 
mixing terms are defined in the text.}
\label{fig-l}
\end{figure}

Where are we headed? Employing the particle content described above as well as the Higgs scalars that we will soon encounter, we want to obtain a 1-loop Dirac neutrino mass term, \ie,
\begin{equation}
{\cal L}_{Dirac}=m_\nu \bar \nu_R\nu_L+h.c.\,.
\end{equation}
Such a term can be seen to be generated via a 1-loop diagram of the form shown in Fig,~\ref{fig-l} whose various component elements we will need to assemble, some of which will naturally be part 
of any $E_6$-like setup but others we will require to included here by construction. These various pieces we will discuss in turn as we go through the set of Higgs scalars that we already require 
for charged fermion mass generation at the appropriate mass scales. As noted above, for simplicity we will only consider the case of a single generation but the analysis we present can be easily 
generalized to the more realistic case.

The simplest fermion mass to generate is the case of the SM $u$-quark, as neither of its chiral components carries any $2_I1_{Y_I}$ quantum numbers and so is a $G_D$ singlet. A 
conventional, $Z_2$-even,  SM-like Higgs doublet $H =(H^+,H^0)^T \sim (1,2,1/2,1,0)$ can thus generate the required mass term via the usual SM-like coupling 
\begin{equation}
y_u \bar u_R \begin{pmatrix}u_L \\ d_L \\ \end{pmatrix}_i \begin{pmatrix} H^+ \\ H^0 \\ \end{pmatrix}_j \epsilon^{ij} +h.c.\,,
\end{equation}
when $H^0$ obtains a vev $\left<H^0 \right> =v/\sqrt 2$ with $v$ at the electroweak scale, $v\sim 100$ GeV.  Now on the other-hand, for the $d$-quark, a different $Z_2$-even Higgs multiplet, $B$, is 
required, with a Yukawa coupling of the form
\begin{equation}
y_d \begin{pmatrix} \bar D_R & \bar d_R \\ \end{pmatrix}_J \begin{pmatrix}u_L \\ d_L \\ \end{pmatrix}_i \begin{pmatrix} h_a^0 & h_b^0 \\ h_a^- & h_b^- \\  \end{pmatrix}_{jJ} \epsilon^{ij} +h.c.\,,
\end{equation}
from which it is clear that $B$ is a $2_L2_I$ bi-doublet, \ie, $B \sim (1,2,-1/2,2,1/2)$. Here we have used the convention that $SU(2)_L$ lower-case indices, $i,j$, label rows (since it acts 
vertically) while $SU(2)_I$ upper-case indices $I,J$ label columns (as it acts horizontally). Note that since both of the $T_{3L}=1/2$ entries in $B$ are electrically neutral, $h_{a,b}^0$, both of these 
fields may obtain vevs, $v_{a,b}/\sqrt 2$, with $v_b \sim 100$ GeV. Since 
 $Q_D(h_{a,b}^0)=1,0$, $v_b$ (in combination with $v$ above) breaks the SM gauge group but will have no impact on $U(1)_D$ breaking. On the other hand, since $h_a^0$ carries a dark charge 
 its vev generates a mass for the dark photon and so we require that $v_a \lsim$ 1 GeV or so. Thus we see that while $v_b$ generates the usual $d$ mass, $v_a\neq 0$ generates mass-mixing 
 between $d$ and $D$, so that $D$ becomes unstable, a role that is generically filled by the dark Higgs. Furthermore, we note that since $h_a^0$ has {\it both} $T_{3L}$ and $Q_D \neq 0$ 
 it will also generate a small mass mixing between the dark photon and the SM $Z$ which is not suppressed by a KM factor, $\epsilon$, but instead only by the square of the vev hierarchy ratio 
 which, in the end, is not so numerically different from $\epsilon$ in magnitude, \ie,  $\sim 10^{-4}$ or so, and may even be somewhat smaller in magnitude. There are, however, good reasons to 
 expect that the contribution of this vev to the DP mass is suppressed relative to the others that we will encounter below\cite{next}.

 The corresponding Yukawa coupling for the electron to $B$, \ie, 
\begin{equation}
y_e \bar e_R \begin{pmatrix} \nu_L & N_L \\ e_L & E_L \\ \end{pmatrix}_{iI} \begin{pmatrix} h_a^0 & h_b^0 \\ h_a^- & h_b^- \\  \end{pmatrix}_{jJ} \epsilon^{ij} \epsilon^{IJ}+h.c.\,,
\end{equation}
is also seen to generate the electron mass as well as an $e-E$ mass mixing via this same small vev, $v_a$, as in the $Q=-1/3$ quark sector but with opposite apparent helicity; this mixing then 
allows $E$ to decay.

A different $Z_2$-even Higgs field is needed to generate the diagonal heavy PM charged fermion ($D,E$) masses (note that $m_N=m_E$) via the couplings
\begin{equation} 
y_D\begin{pmatrix} \bar D_R & \bar d_R \\ \end{pmatrix}_I D_L \begin{pmatrix} \phi_{1,a}^0 & \phi_{1,b}^0 \\  \end{pmatrix}_J \epsilon^{IJ}+h.c.\,, 
\end{equation}
and
\begin{equation}
y_E \begin{pmatrix} \bar N_R \\ \bar E_R \\ \end{pmatrix}_i \begin{pmatrix} \nu_L & N_L \\ e_L & E_L \\ \end{pmatrix}_{iI} \begin{pmatrix} \phi_{1,a}^0 & \phi_{1,b}^0 \\  \end{pmatrix}_I  +h.c.\,,
\end{equation}
where $\Phi_1 = (\phi_{1,a}^0,\phi_{1,b}^0) \sim (1,1,0,2,1/2)$, with both neutral members obtaining vevs, $(v_1,V_1)/\sqrt 2$. Here we see that $Q_D(\phi_{1,b}^0)=0$ so that 
$V_1 \sim 10$ TeV generates the large $N,E$ and $D$ Dirac masses while simultaneously breaking $2_I1_{Y_I} \to 1_D$, thus contributing to the masses of both the $W_I^{(\dagger)}$ 
and $Z_I$ gauge 
bosons. From direct searches for such heavy states at the LHC we know that the masses of $N,E,D$ all roughly lie above $\sim 1-1.5$ TeV\cite{Rizzo:2022qan}.  
Since $Q_D(\phi_{1,a}^0)\neq 0$, the vev $v_1 \sim $ 1 GeV or so also contributes to the breaking of $1_D$ but does {\it not} generate any additional $Z$-dark photon mass mixing since it is a $2_L$ singlet. It also leads to 
a further contribution to both the $e-E$ and $d-D$ mass mixings (but of the opposite apparent helicity) at the $\sim $ GeV scale.  We also see that the $y_E$ Yukawa coupling generates the necessary 
interaction,  $\bar N_R \nu_L\phi_{1,a}$, appearing as the first step in the Dirac neutrino mass generation diagram as shown in Fig.~\ref{fig-l} going from left to right in the loop.

\subsection{Additional Scalar Couplings}

Finally, we come to the couplings involving the SM singlet fields $S_{L,R}$ and $\nu_R$. Here we recall that since we've assigned $\nu_R$ to have a general dark $U(1)_D$ charge, 
$Q_D(\nu_R)=q\neq 0$, it must be, by $2_I$ invariance, that $Q_D(S_R)=q-1=Q_D(S_L)$ since PM(-like) fields generally couple to $U(1)_D$, \ie, the DP, in a vector-like manner. Hence, the first 
such coupling, which is responsible for the next step in the generation of the $\nu$ Dirac mass term, takes the form
\begin{equation}
y_S  \begin{pmatrix} \bar N_R \\  \bar E_R \\ \end{pmatrix}_i S_L\begin{pmatrix} H'^+ \\ H'^0 \\ \end{pmatrix}_j \epsilon^{ij}+h.c.\,,
\end{equation}
where the $Z_2$-even Higgs doublet, $H'$, must now transform as $H' =(H'^+,H'^0)^T \sim (1,2,1/2,1,-q)$ since $Q_D(N_R)=-1$. When $H'^0$ obtains a vev $\left<H'^0 \right> =v'/\sqrt 2$ with $v'$ at 
the $U(1)_D$- breaking scale, $v'\lsim 1$ GeV, as it must be since it carries $Q_D\neq 0$, a mass mixing of the form $\sim y_S v' \bar{S_L}N_R$ is thereby generated as is shown in Fig.~\ref{fig-l}. 
As in the $B$ case above, the small $v'$ vev will, besides contributing to the familiar SM $W,Z$ masses will also induce a mass mixing between the $Z$ and the DP. 

The last pair of Yukawa couplings that we need will generate both the large $\lsim 10$ TeV Dirac mass for $S$ as well as the required $\bar \nu_R S_L\phi_{2,a}$ coupling seen in Fig.~\ref{fig-l} and 
are given by
\begin{equation}
 \begin{pmatrix}  \bar \nu_R & \bar S_R \\ \end{pmatrix}_I S_L \Big(y_R\begin{pmatrix} \phi_{2,a}^0 & \phi_{2,b}^0 \\ \end{pmatrix}_I+y_R'\begin{pmatrix} \phi_{2,a}^0 & \phi_{2,b}^0 \\ \end{pmatrix}'_I\Big)+ h.c.\,,
\end{equation}
where, since the $2_I$ doublet $(\nu_R,S_R)$ is $Z_2$-odd, the corresponding $Z_2$-odd Higgs fields are just $\Phi_2^{(')} = (\phi_{2,a}^0,\phi_{2,b}^0)^{(')} \sim (1,1,0,2,1/2)$, with one upper 
and one lower neutral member obtaining the vevs $(v_2,0)/\sqrt 2$ and $(0,V_2)/\sqrt 2$, respectively. As $Q_D(\phi_{2,a},\phi_{2,b})^{(')}=(1,0)$, we expect that $V_2\sim 10$ TeV, contributing to 
the $2_I1_{Y_I}$ symmetry breaking, as does $V_1$ above, 
and generating the large mass for $S$, while the $U(1)_D$-breaking $v_2\sim 1$ GeV allows for the decay $S\to \nu +$DP via mixing, similar to what we've encountered for the more familiar $N,E$ 
and $D$ PM fermions above. This coupling with $\phi_{2a}$ also provides another missing piece in Fig.~\ref{fig-l}. 
Note that now $\phi_{1a}$ and $\phi_{2a}$, which are associated with the small vevs $v_{1,2}$, will at least partially play the role of the usual dark Higgs field which breaks 
$U(1)_D$, so we might expect that their masses may also be {\it relatively} small ( based on unitarity arguments) or at least that they will easily satisfy $m_{1,2}^2 << m_{N,E,D,S}^2$, the latter 
being generated by the 
$\sim 10$ TeV scale vevs $V_{1,2}$, while all of the other new scalar fields transforming nontrivially under the dark gauge group, $G_D$, are assumed to be rather heavy, associated with this 
high mass scale. We will make use of this expectation in our numerical analysis below.

We note that the above set of six Higgs fields that we have introduced, which are summarized in Table \ref{higgstab}, leads to a very extensive scalar potential, even after the imposition of both the 
gauge and $Z_2$ discrete symmetries, some important terms of which for our discussion below take the form 
\begin{equation}
\begin{aligned}
U= &(\Phi_1^\dagger \Phi_1)\Big[\lambda_1 (\Phi_2^\dagger \Phi_2) +\lambda_2({\Phi'_2}^\dagger \Phi'_2)\Big]+\lambda_3({\Phi'_2}^\dagger \Phi'_2)({\Phi_2}^\dagger \Phi_2)
+\lambda_4(\Phi_1^\dagger \Phi_2)(\Phi_2^\dagger \Phi_1)\\& +\lambda_5(\Phi_1^\dagger \Phi'_2)({\Phi'_2}^\dagger \Phi_1)+\lambda_6({\Phi'_2}^\dagger \Phi_2)(\Phi_2^\dagger \Phi'_2) 
 +...  ~. 
\end{aligned}
\end{equation}
Now the last ingredient we see in Fig.~\ref{fig-l} that is needed to complete the diagram is to generate a $\phi_{2,a}-\phi_{1,a}$ mass mixing term. Enforcing both $G_{SM}\times G_D$ and 
$Z_2$ symmetries on the scalar potential as above implies that, \eg, a possible $\sim \mu_{12} \Phi_1^\dagger \Phi_2 +h.c$ term, which necessarily violates both the $Z_2$ symmetry as well 
as $U(1)_D$,  is obviously excluded. Thus the only way to generate such a scalar mixing term is via a pair of the allowed quartic terms above involving both $\Phi_{1,2}$, in particular, the 
ones that are proportional to $\lambda_1$ and $\lambda_4$ (which are assumed to both be of order unity) 
and then to extract the terms proportional to the product of the two $\sim1$ GeV vevs, \ie, $m_{12}^2=(\lambda_1+\lambda_4) v_1v_2 \lsim 1 ~{\rm GeV}^2$, and 
which is seen to also proportional to required field combination $\phi_{2,a}^\dagger \phi_{1,a}+h.c.$. 

Lastly, as we noted, that since additional SM-singlet dark sector fermions can be (and must be to explain DM and to cancel anomalies) added to this setup, any additional corresponding scalars that are 
required can also easily be included in the structure above. 

\begin{table}
\begin{center}
\caption{Minimal Higgs sector content for the present setup - additional SM singlet dark sector Higgs fields may also exist.}\label{higgstab}
\begin{tabular}{ l c c c c c c}
\hline
  & SU(2)$_L$ & $Y/2$ & SU(2)$_I$ & $Y_I/2$ & $Z_2$ & vev(s) \\
\hline
$H$ & {\bf 2} & 1/2 & {\bf 1} & 0 & +& $\frac{1}{\sqrt{2}}\begin{pmatrix} 0 \\ v \\ \end{pmatrix} \sim \begin{pmatrix} 0 \\ 100 \textrm{ GeV} \\ \end{pmatrix}$  \vspace{0.1cm}\\
$H'$ & {\bf 2} & 1/2 & {\bf 1} & -q & + &  $\frac{1}{\sqrt{2}}\begin{pmatrix} 0 \\ v' \\ \end{pmatrix} \sim \begin{pmatrix} 0 \\ 1 \textrm{ GeV} \\ \end{pmatrix}$  \vspace{0.1cm}\\
$B$ & {\bf 2} & -1/2 & {\bf 2} & 1/2 & + &  $\frac{1}{\sqrt{2}}\begin{pmatrix} v_a & v_b \\ 0 & 0 \\ \end{pmatrix} \sim \begin{pmatrix} 1 \textrm{ GeV} & 100 \textrm{ GeV} \\ 0 & 0 \\ \end{pmatrix}$ \vspace{0.1cm}\\
$\Phi_1$ & {\bf 1} & 0 & {\bf 2} & 1/2 & + & $\frac{1}{\sqrt{2}}\begin{pmatrix} v_1 & V_1 \\ \end{pmatrix} \sim \begin{pmatrix} 1 \textrm{ GeV} & 10 \textrm{ TeV} \\ \end{pmatrix}$ \\
$\Phi_2$ & {\bf 1} & 0 & {\bf 2} &  1/2 & - & $\frac{1}{\sqrt{2}}\begin{pmatrix} v_2 & 0 \\ \end{pmatrix} \sim \begin{pmatrix} 1 \textrm{ GeV} & 0 \\ \end{pmatrix}$ \\
$\Phi_2'$ & {\bf 1} & 0 & {\bf 2} &  1/2 & - & $\frac{1}{\sqrt{2}}\begin{pmatrix}  0 & V_2 \\ \end{pmatrix} \sim \begin{pmatrix} 0 & 10 \textrm{ TeV} \\ \end{pmatrix}$ \\
\hline 
\end{tabular}
\end{center}
\end{table}   

\section{Dirac Mass Calculation}

As a first step in discussing the neutrino mass, it is easy to convince oneself that the above limited set of couplings do not allow for a Dirac term to be generated at tree-level in the current setup. To that 
end, we consider the set of tree level neutral fermion Dirac mass terms generated by the vevs of the various Higgs fields introduced in the last Section in the ${\cal N}=(\nu,N,S)$ basis, \ie,  
\begin{equation}
{\cal L}_{mass}= {\cal {\bar N}}_R M {\cal N}_L +h.c.\,,
\end{equation}
with the $3\times 3$ mass matrix $M$ given by  
\begin{equation}
M=\begin{pmatrix} 0 & 0 & y_Rv_2 \\ y_Ev_1 & m_N & y_Sv' \\0 & 0& m_S \\ \end{pmatrix}\,,
\end{equation}
with $m_{N,S}$ being the $N,S$ masses, lying at the TeV scale as discussed above, and the off-diagonal entries generated  by the various $U(1)_D$-breaking vevs lying $\lsim 1$ GeV. For simplicity, 
we define  $a=y_Rv_2,b=y_Ev_1$ and $c=y_Sv'$, where clearly we have that $a^2,b^2,c^2<< m_{N,S}^2$. We then obtain that 
\begin{equation}
MM^\dagger=\begin{pmatrix} a^2 & ac & a m_S\\ ac & m_N^2+b^2+c^2 & cm_S \\am_S & cm_S& m_S^2 \\ \end{pmatrix}\,,
\end{equation}
from which we find immediately that $det (MM^\dagger)=0$ having only two non-zero eigenvalues, $\simeq m_{N,S}^2$, so that the SM neutrino remains massless. Thus in this setup, a Dirac neutrino 
mass does not occur at tree-level and can only potentially arise at the 1-loop level or higher.

\begin{figure}[htbp] 
\centerline{\includegraphics[width=5.0in,angle=0]{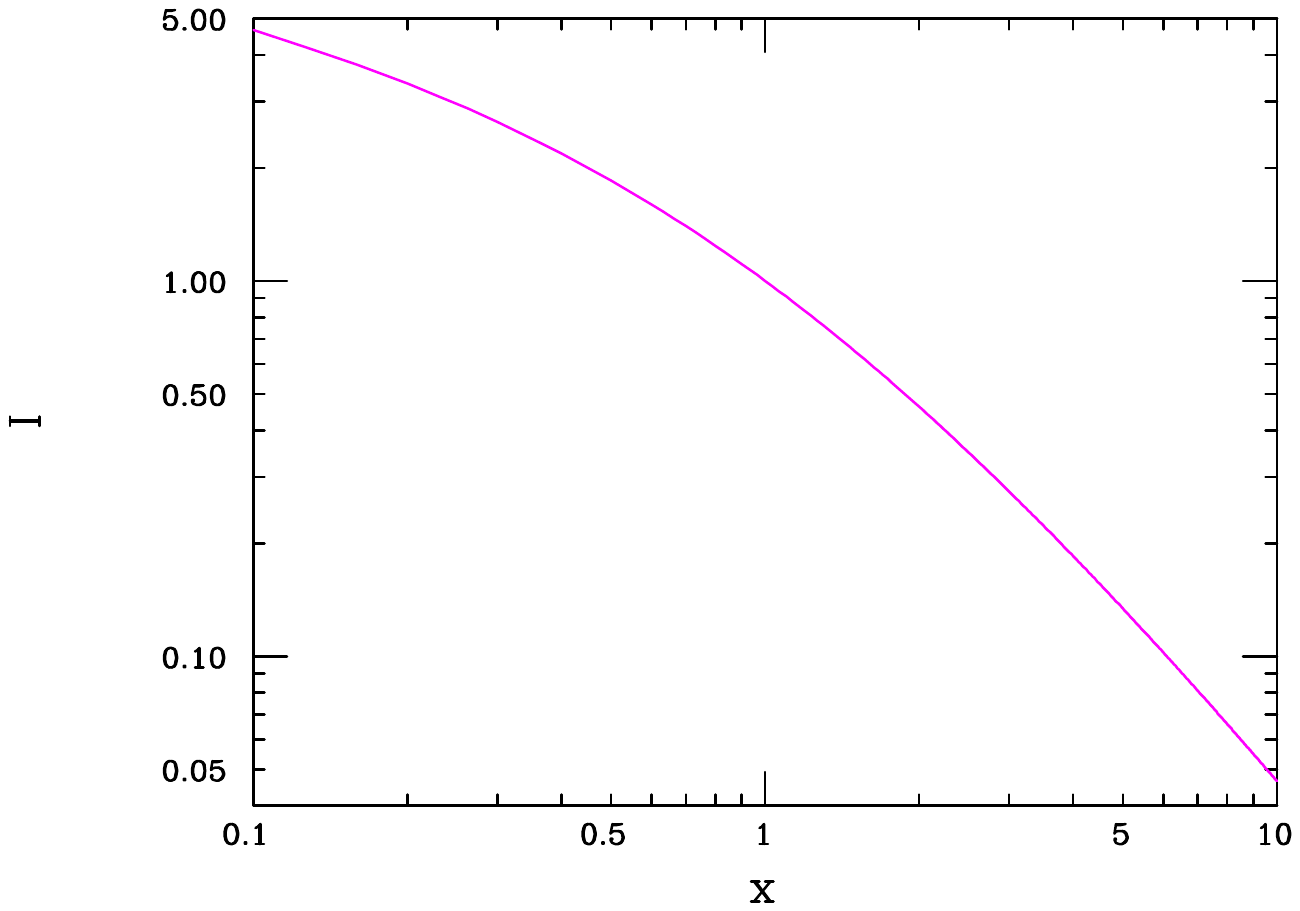}}
\vspace*{-1.3cm}
\caption{The function $I(x)$ as defined in the text with $x$ being the mass ratio $x=m_S/m_N$.}
\label{fig0}
\end{figure}

The needed 1-loop induced Dirac neutrino mass as shown in Fig.~\ref{fig-l}, can be now evaluated explicitly in terms of the product of the three Yukawa couplings $y_{R,E,S}$ introduced above, 
the sum of the quartic couplings, $\lambda_m=\lambda_1+\lambda_4$, the product of the three small $U(1)_D$-breaking vevs $v',v_{1,2}$ and the masses of the fields appearing in the loop, \ie, 
$N,S, \phi_{1a}(=m_1)$ and $\phi_{2a}(=m_2)$ as 
\begin{equation}
m_\nu=\frac{\lambda_m~(y_Ry_Ey_S)~(v'v_1v_2)}{16\sqrt 2 \pi^2}~ J(m_N^2,m_S^2,m_1^2,m_2^2)\,,
\end{equation}
where the function $J$ results from performing the usual parametric integration over the loop momenta, \etc. To get some numerical feel for this result, we will assume for purposes 
of demonstration the following set of suggestive values: $\lambda_m=1/2$, the product of the three Yukawa couplings is $y_Ry_Ey_S=0.1$, the small vevs share a common value, $v',v_{1,2}=1$ GeV 
and we will further assume for simplicity and their roles associated with the dark Higgs that $m_{1,2}^2 << m_{N,S}^2${\footnote {Note that these inequalities will be satisfied even if $m_{1,2}$ lie at 
or the electroweak scale.}} as was mentioned above. In this limit, we can then write that 
\begin{equation}
J(m_N^2,m_S^2,m_{1,2}^2/m_{S,N}^2\to 0)\simeq \frac{1}{m_N^2}~I(x)\,,
\end{equation}
where we have defined the mass ratio $x=m_S/m_N$, which we expect to be not too far from unity, and where the function $I(x)$ is explicitly given by the rather simple expression  
\begin{equation}
I(x)=\frac{2~ {\rm {log}}(x)}{x^2-1}\,.
\end{equation}
We then obtain that the Dirac neutrino mass is given numerically by  
\begin{equation}
m_\nu \simeq 0.056~{\rm eV}\cdot \Big(\frac{\lambda_m}{0.5}\Big)~\Big(\frac{y_Ry_Ey_S}{0.1}\Big)~\Big(\frac{v'v_1v_2}{1~{\rm GeV}^3}\Big)~\Big(\frac{2 ~{\rm TeV}}{m_N}\Big)^2 ~I(x)\,,
\end{equation}
with the values of the function $I(x)$ for $0.1 \leq x \leq 10$ now appearing in Fig.~\ref{fig0} where we see that $I(x)>(<)1$ when $x<(>)1$ and can vary away from unity by roughly an order of 
magnitude.  Clearly this simple numerical exercise suggests that Dirac 
neutrino masses in the desired regime near $\sim 0.05$ eV can be easily obtained over a respectable correlated set of ranges for the various model parameters appearing above.

\section{Model Tests at Colliders}

The possibility that neutrinos are Dirac particles whose right-handed components carry dark quantum numbers is difficult to test, either directly or even indirectly,  in terrestrial experiments 
as conventional sources produce neutrinos with MeV scale energies or above so that the neutrinos are always very highly boosted. This combined with the fact that the SM neutrino interactions 
are universally left-handed implies that the additional interactions associated with this usual nature for the neutrino would typically be extremely helicity suppressed and so quite difficult to observe 
at best. This possibility, as well as the potential astrophysical and cosmological implications of such a scenario arising from, \eg, the CMB and BBN, that more than likely may yield even stronger 
constraints in some regions of the model parameter space{\footnote {Recall, however, that the usual left-handed neutrinos do not couple to dark photons as they have both $Q_{em}=0$ and $Q_D=0$ 
unlike, \eg, in the case of a light $U_{B-L}$ gauge boson. Here, in contrast to much of the literature, only the right-handed neutrinos experience these new gauge interactions.} than 
do colliders\cite{astro}, will be discussed in future work\cite{elsewhere}. 

In addition to predicting that neutrinos are Dirac fermions, the real tests of this setup may then be to discover its various building blocks and to examine the properties of these various new states that 
it predicts. In almost all cases, since the new fermionic states are heavy, this will require a high energy collider. Of course, given the $E_6$-like nature of this scenario, many of the features of 
the present setup are shared by a wider class of models which will necessarily display many common signal elements, \eg, the PM fields themselves with their predicted production and decay 
properties as well as the new heavy gauge bosons and Higgs fields. Production of $S_{L,R}$ will, of course, be more problematic as it is totally decoupled from the SM at tree-level; ideally, we'd like to 
have some access to this new heavy state. 

Since the particles participating in the loop diagram generating the Dirac neutrino masses in Fig.~\ref{fig-l} are all electrically neutral and are also mostly SM singlets, the direct signatures for 
this model, \ie, the production and observation of the relevant new fermion and Higgs field degrees of freedom at colliders is, to say the least, non-trivial. Of the newly introduced color singlet 
fermionic PM states, only $N_{L,R}$, which occurs in an $2_L$ (bi)doublet with the corresponding $Q_{em}=-1$ states, $E_{L,R}$, directly couple to SM fields via conventional the $W,Z$-exchange 
interactions. Dark Higgs fields, which do not have SM interactions, may be produced in PM decays or via $s$-channel SM Higgs exchange provided that the appropriate quartic couplings from 
which these 
interactions arise are reasonably large and these dark states are not overly massive as has been discussed in earlier work\cite{Rizzo:2026dsj}. However, the decays of such particles are likely to 
also be dominated by lighter dark sector final states rendering their production and subsequent decay invisible, \ie, simply producing just missing $E_T$, if they occur without any of the usual 
additional accompanying 
conventional tagging particles, \eg, a jet, a $W/Z$ or a SM Higgs boson. Further, at least some of the relevant quartic couplings arising from the potential may also already be required to be suppressed 
based on the existing constraints on the invisible decay branching fraction for the 125 GeV Higgs as determined at the LHC\cite{Invisible}.

\begin{figure}[htbp] 
\centerline{\includegraphics[width=5.0in,angle=0]{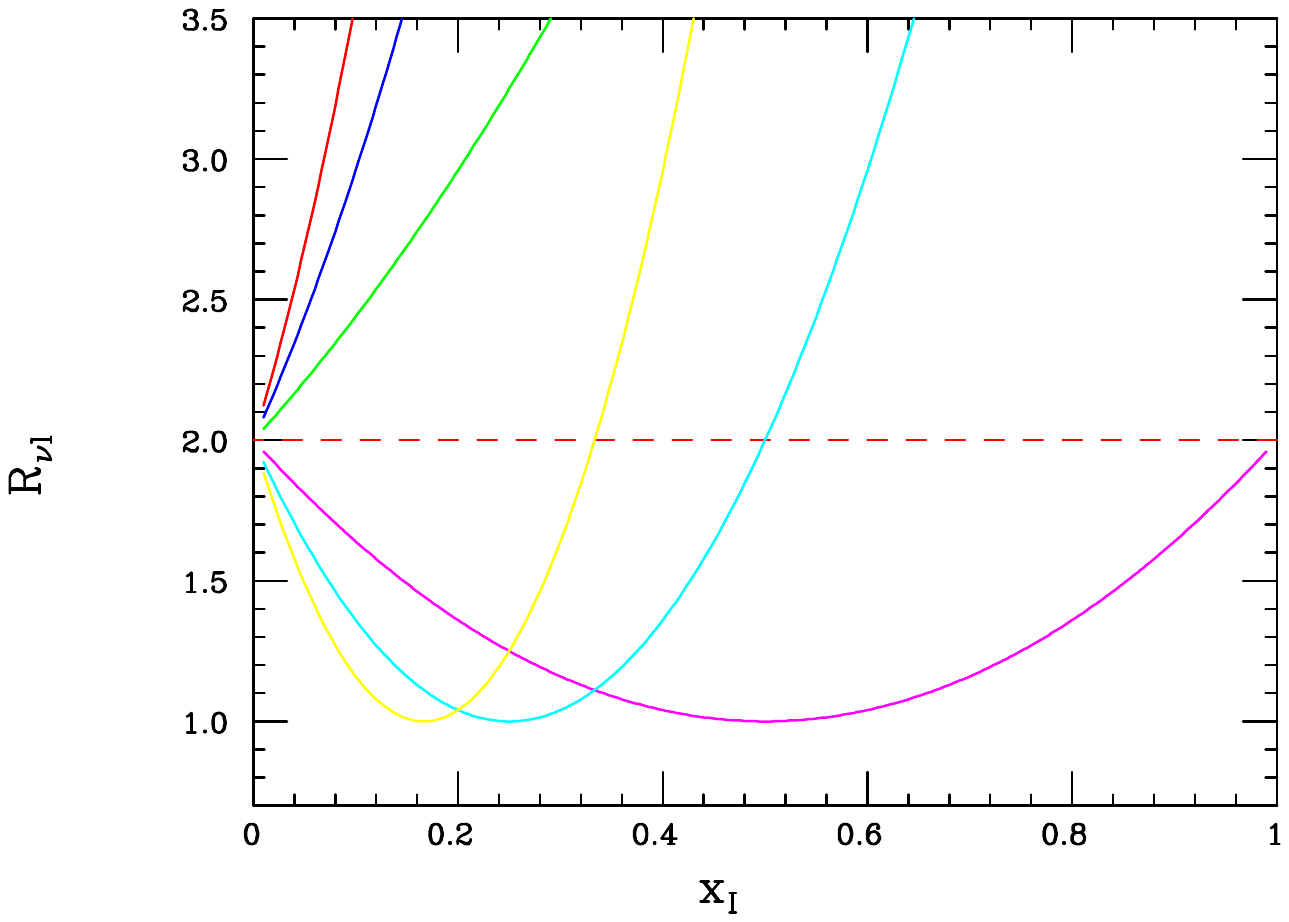}}
\vspace*{-1.3cm}
\caption{The ratio $R_{\nu l}$ for a single generation, as defined in the text, as a function of $x_I$ assuming that $q=-3,-2,-1,1,2,3$ corresponding to the red, blue, green, magenta, cyan and 
yellow curves, respectively. The red dashed line corresponds to the familiar $q=0$ Dirac or the Majorana seesaw expectation, $R_{\nu l}=2$, for purposes of comparison.}
\label{figrnl}
\end{figure}

{\it In principle}, the production and the decay of the heavy neutral $G_D$ analog of the SM $Z$ boson, \ie, the $Z_I$, might offer a direct window into whether or not $\nu_R$ carries a non-zero 
dark charge, $Q_D=q$, which 
is a feature of the current setup. The basic idea is rather simple: measure the ratio of the $Z_I$  branching fractions/decay widths into $\bar \nu \nu$ and $\ell^+ \ell^-$ where 
$\ell=e,\mu$ or $\tau$, assuming generational universality. That a measurement of this kind is possible and can even be performed with rather high precision ($\sim 2-3\%$) at a hadron collider has 
been shown to be the case by both the ATLAS\cite{ATLAS:2023ynf} and CMS\cite{CMS:2022ett} collaborations employing $\sqrt s=13$ TeV data at the LHC for the case of the SM $Z$ as will 
be discussed further below. Why is such a ratio useful? First, if the $Z_I$ can {\it only} decay to SM final 
states because the PM fields are too massive, \ie, $2m_{N,E,D,S}>m_{Z_I}$ in a fashion analogous to the top quark and the SM $Z$, then 
$N_g\Gamma(Z_I\to \bar \nu \nu)=\Gamma(Z_I\to ~invisible)$, where $N_g=3$ is the number of generations. 
Second the ratio of these leptonic partial widths is just given by (in the present, single-generation, example where $\nu_R$ has $T_{3I}=1/2$ with $Q_D=q$)
\begin{equation}
R_{\nu l}=\frac{\Gamma(Z_I \to \bar \nu \nu)}{\Gamma(Z_I\to \ell^+\ell^-)}=1+(1-2x_Iq)^2\,,
\end{equation}
recalling that $0<x_I <1$ is just the analog of $x_w=s_w^2$ in the SM, \ie, $g_D^2=g_I^2x_I$ and $g_{Y_I}^2/g_I^2=x_I/(1-x_I)$. However, by contrast, if the neutrino were to be a $q=0$ Dirac or a 
Majorana fermion with $\nu_L$ being light and $\nu_R$ being very massive as in the usual seesaw model then one obtains $R_{\nu l}=2$. Fig.~\ref{figrnl} shows this ratio as a function of $x_I$ 
for several different (here assumed to be integer) values of $q$; from this we see that even a modest 
determination of this ratio at the, say, $\sim 20-30 \%$ level, would clearly distinguish the present scenario from the familiar $q=0$ Dirac or Majorana neutrino seesaw setups over a large part of the 
model parameter space. For example, if $x_I \simeq s_w^2=x_w$, we see that all of the $q\neq 0$ predictions lie quite far from the more `conventional' expectation. 

As the observation of the dilepton final state of the $Z_I$ is `straightforward' (assuming that the $Z_I$ is kinematically accessible), and is the conventional discovery channel, the essential problem in 
making this ratio measurement is that it would also involve observing the decay to an invisible final state from the neutrinos, \ie, appearing as just $E_T^{miss}$. That problem has been solved, at 
least in principle, as noted above by both 
ATLAS\cite{ATLAS:2023ynf} and CMS\cite{CMS:2022ett}, by the use of high-$p_T$ jet tagging, \ie, requiring that both the dilepton and neutrino pairs (\ie, MET) to be 
always accompanied by at least one high-$p_T$ jet. Then the ratio of the monojet and the dilepton+jet $p_T$ distributions, after background subtractions and corrections for differences in acceptances 
and efficiencies, would yield the value of $R_{\nu l}$. Clearly this requires that these backgrounds be quite well understood and that reasonable statistically significant signal 
samples would be available as was the case for the 
measurements of the SM $Z$. In the present case, we face some immediate obstacles to this program $(i)$ as the $Z_I$ itself has not (yet) been discovered which for SM strength couplings, 
\ie, $g_I/c_I=g/c_w$,  implies that $m_{Z_I}\gsim 5$ TeV\cite{Rueter:2019wdf,Rizzo:2022qan}. Here, we remind ourselves that the $Z_I$ does not couple to the $u-$quark and, since SM fermions all 
have $Q_D=0$, their couplings to the $Z_I$ are explicitly independent of $x_I$. This implies that if the $Z_I$ has only SM particle final states in its decays (ignoring $\nu_R$ for the moment) then the 
search reach will be set only the magnitude of the overall coupling $g_I/c_I$ relative to, say, the corresponding SM quantity (this ratio called called $r$ in our previous work) at least in the narrow 
width approximation. $(ii)$ Note that relatively large samples (certainly $\gsim 100$ events of both types) of $Z_I+$jet events are required to make the 
measurement with reasonable precision. Given that the planned total integrated luminosity of the HL-LHC is $\sim 3$ ab$^{-1}$ so that roughly $\sim 10x$ as much data as we have at present 
will eventually be available, it seems very unlikely that this measurement could be made there - even if the $Z_I$ were to be soon discovered - due to the poor statistics. This would mean that we would 
need to wait, on the hadron collider side, for the much higher energy (and higher integrated luminosity) FCC-hh.  We note, however, that even there the possibility of the on-shell production 
of, \eg, the $N$ and/or $S$ states in 
$Z_I$ decay and their subsequent decay to SM neutrinos plus dark photons would certainly lead to a substantial contamination of this signal, likely rendering a significantly precise measurement 
of $R_{\nu l}$ extremely difficult. However, if a $\sim 10$ TeV lepton collider were to be available on roughly the same time scale as FCC-hh, the measurement of $R_{\nu l}$ would be 
relatively straightforward following the same approach as that employed at both LEP and the SLC since the $Z_I$ would hopefully be able to be produced on-shell at such a machine.

Of course, even if the $Z_I$ is lighter than (m)any of the PM states, its virtual $s-$channel exchange can lead to the production of both $S$ and $\nu_R$ with respectable rates. However, the 
difficulty is, again, that their decays, \eg, $S\to \nu_R\phi_{2,a}$, will likely just yield more invisible final states which are difficult to detect.

\begin{figure}[htbp]
\centerline{\includegraphics[width=5.0in,angle=0]{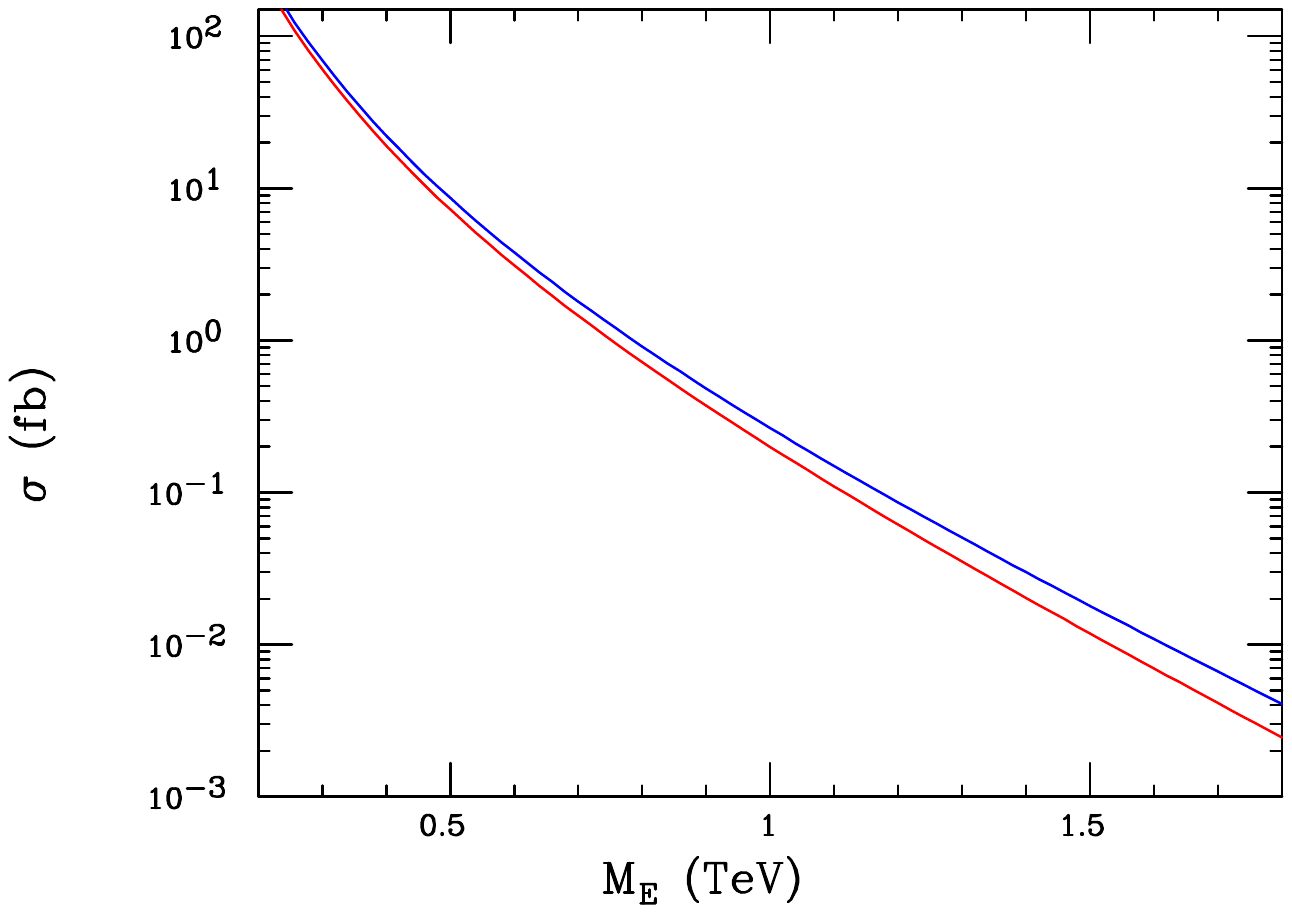}}
\vspace*{-1.2cm}
\centerline{\includegraphics[width=5.0in,angle=0]{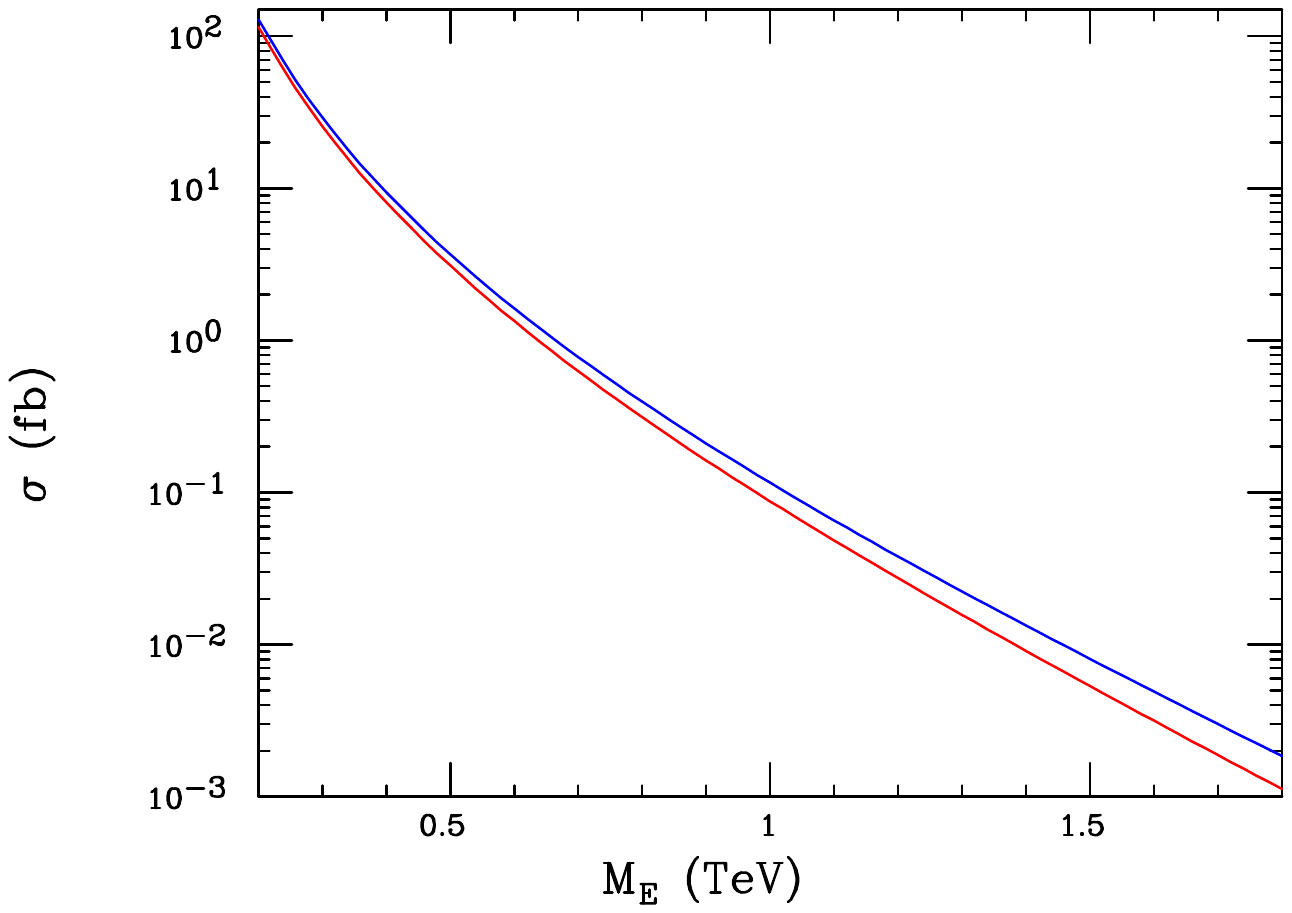}}
\vspace*{-1.3cm}
\caption{Cross sections for vector-like isodoublet (top) and isosinglet (bottom) $Q_{em}=-1$, PM pair production, $E^+E^-$, as a function of mass via $s$-channel SM $\gamma,Z$ exchange 
(with $Z_I$ decoupled) at the LHC assuming $\sqrt s=13$ (bottom red) or 14 (top blue) TeV.}
\label{fig1}
\end{figure}

The other $2_I$ gauge boson, $W_I$, will not help us much; it connects the fields sharing (here) an isodoublet and can mediate the associated production of new PM states which share this 
doublet with a SM partner, \eg, $gd\to W_ID$ and/or $\gamma e \to W_IE$. However, for $2_I$ singlets or for isodoublets {\it not} having a SM member, such as $(\nu_R,S_R)$ in the present case, 
$W_I$ exchange is clearly not useful to access these new states.

What other discovery channels are there that can probe this setup and may be accessible at the HL-LHC or at a future hadron collider such as FCC-hh? 
A necessary ingredient of the current setup is the existence of the color-singlet vector-like 
PM fields, $N,E$, which form a $2_L$ doublet and so can be produced via the $s$-channel exchanges of the SM $W,Z/\gamma$ gauge bosons (independently of the existence or mass of the $Z_I$, 
for example) similarly to the familiar Drell-Yan process. Observing such heavy VL PM states is, however, only a {\it necessary} prediction of the current setup but does not, in itself, {\it guarantee} 
that the present model under discussion is the correct one since these PM states are common to all $E_6-$like setups. Hence, here we only briefly summarize their accessibility at the LHC and 
FCC-hh. While $Z/\gamma$ (and, in principle, $Z_I$) exchange mediates $E^+E^-$ PM pair production, $W^\pm$ exchange will mediate $\bar NE+h.c.$ associated production. As noted 
in earlier work\cite{Rizzo:2022qan}, since $E^\pm$ dominantly decays to a SM lepton, $\ell^\pm$, plus a dark Higgs or a (longitudinally polarized) DP via the Gauge Boson Equivalence 
Theorem\cite{GBET}, the $E^+E^-$ final state appears as $\ell^+\ell^-$+ missing $E_T$ at a collider such as the LHC so that the usual SUSY slepton searches can be rather straightforwardly 
recast\cite{Guedes:2021oqx}. From Run II data at $\sqrt s=13$ TeV, masses below $\simeq 0.90(1.05)$ TeV for isosinglet (isodoublet) first/second generation PM of this kind have been excluded while 
the HL-LHC is expected to be sensitive out to corresponding masses as large as 1.45(1.65) TeV in the limit that the $Z_I$ is sufficiently massive that its contribution to the cross section 
can be neglected; see Fig.~\ref{fig1} for a comparison of 
these two cross sections. The isosinglet (isodoublet) case is seen to have the smaller (larger) cross section due to destructive (constructive) interference between the SM $\gamma$ and $Z$ 
$s$-channel exchanges above the $Z$ pole. If these states are not too massive so that sufficient statistics are available, the isodoublet case (of interest to us here) could be separated from that 
of the isosinglet just from an extracted production cross section measurement alone. 
To probe higher masses via pair production, the FCC-hh will be required as is shown in Fig.~\ref{fig2} where for $E$ being in an isodoublet masses as large as 
3(4) TeV may be reachable for $\sqrt s=60(100)$ TeV.
The corresponding $\bar NE+h.c.$ final state, which occurs only when the $E$ is in an isodoublet together with $N$ (as is the case here and with which it is highly degenerate) can yield a slightly 
greater production cross 
section due to the larger coupling of the charged currents to the SM $W$ as shown in Fig.~\ref{fig3}. Since $N$ decays to a neutrino plus a dark Higgs or dark photon, this yields the final state 
$\ell^\pm$ + missing $E_T$, similar to that of a heavy $W'$ but without a peak in the transverse mass distribution and thus has a larger background arising from the tail of the SM charged-current 
Drell-Yan process. This implies a search reach for this mode which is roughly comparable to that found for the isodoublet $E^+E^-$ production mode at both the HL-LHC and FCC-hh. A signal 
for this $\ell^\pm+$MET final state {\it simultaneously}  with $\ell^+\ell^-+$MET at the expected rates would at least strongly indicate the existence of the $(N,E)$ doublet PM fermions which is a 
necessary ingredient of the current setup.

\begin{figure}[htbp]
\centerline{\includegraphics[width=5.0in,angle=0]{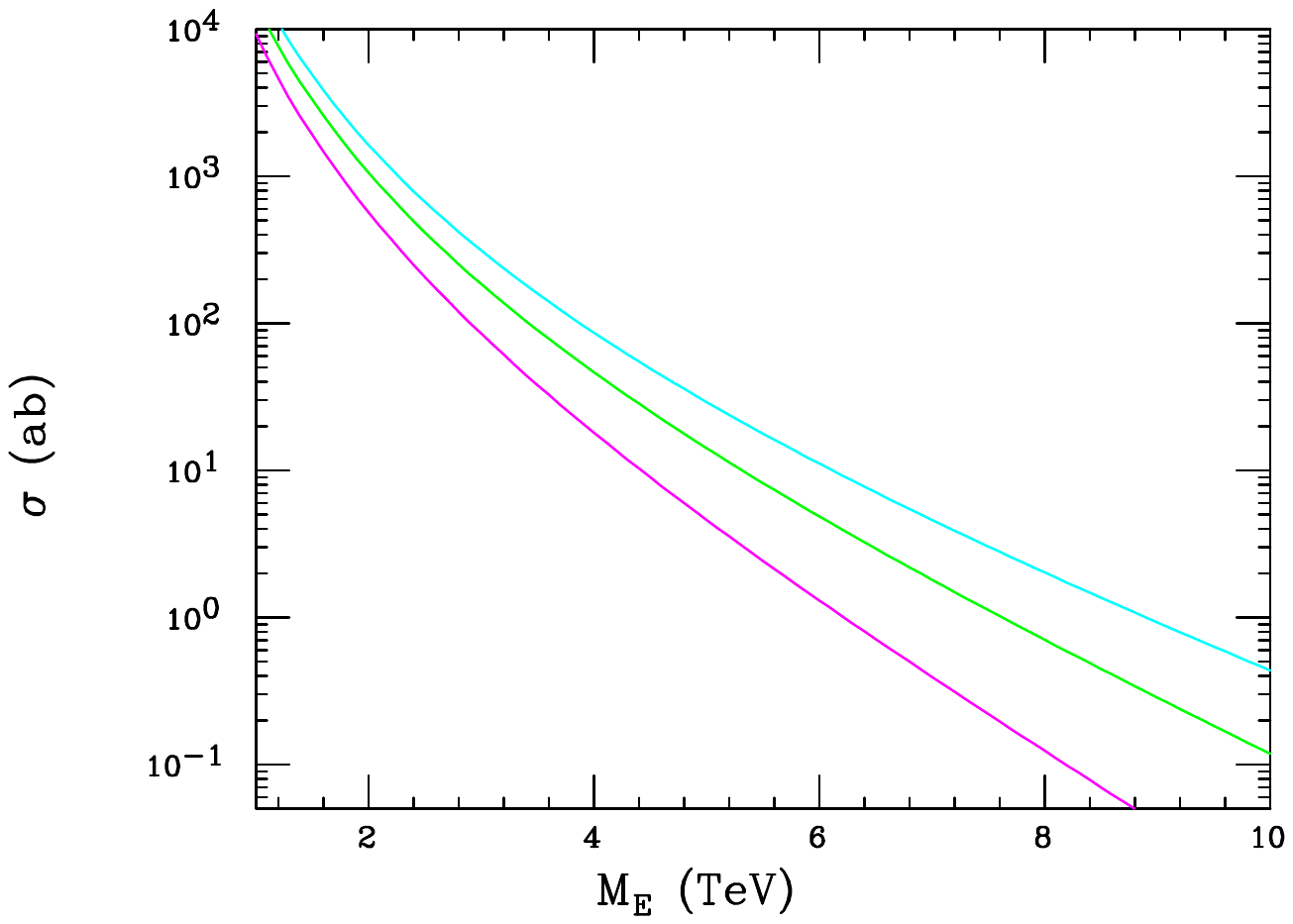}}
\vspace*{-1.2cm}
\centerline{\includegraphics[width=5.0in,angle=0]{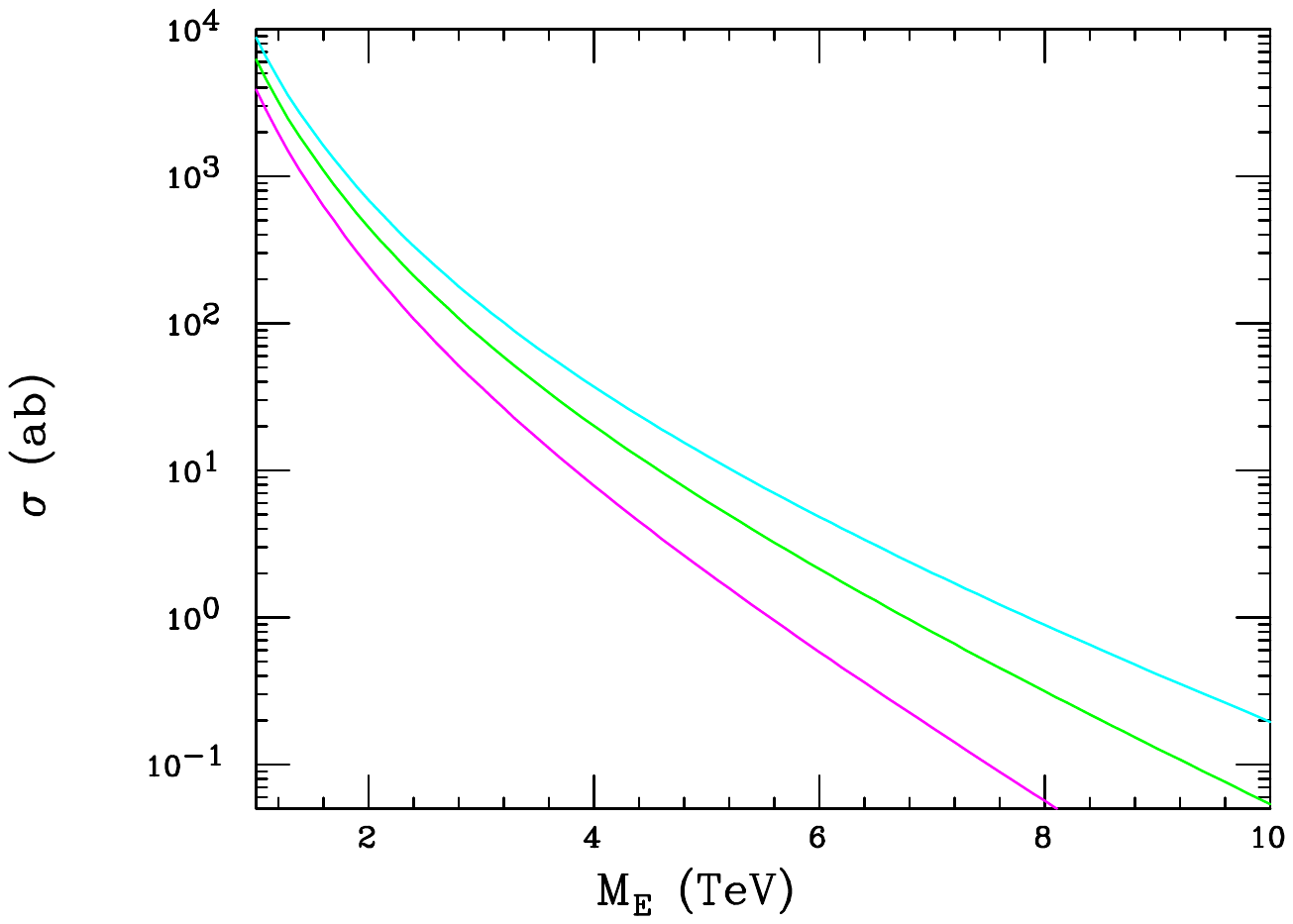}}
\vspace*{-1.3cm}
\caption{Same as in the previous Figure but now for the FCC-hh assuming, from bottom to top, $\sqrt s=$60, 80 or 100 TeV.}
\label{fig2}
\end{figure}

On the other-hand, at a multi-TeV lepton collider, which might be constructed on a similar timescale as FCC-hh, the isodoublet and isosinglet cases for $E^+E^-$ would be easily differentiated by 
both the production cross section as well as the reconstructed angular distribution\cite{Rizzo:2025fsy}. This is possible even if additional $t-$channel dark Higgs exchange contributions are present 
at a significant level which can occur when the initial lepton and the $E$ flavor are the same.

In the case of the heavy neutral VL lepton PM, $N$, pair production, which occurs in the isodoublet case via SM $Z$ exchange with a similar cross section to that of the isodoublet $E$, the 
situation is a bit more difficult. Most, if not all, discussions about the production of similar heavy neutral lepton-like states at colliders focus on their mixing with the SM neutrinos of various flavors 
thus leading to decays such as, \eg, $N \to \ell^\pm W^\mp, ~W^\mp \to 2j$ as may occur in seesaw Majorana mass models\cite{ATLAS:2025qbs}. Here, although $N$ indeed mixes with the SM 
neutrinos, they are not (dominantly) accompanied in the final state by a $W,Z$ or SM Higgs but instead by the DP or a dark Higgs as discussed above. These more `SM-like' decays generally will 
only appear at the $\sim$ few percent level unless the relevant Yukawa coupling would for some reason be significantly suppressed, \eg, $y_E<<1$. Relying on only these highly suppressed modes for 
detecting a signal would very much lead to a deterioration in the search reach for any new neutral lepton-like PM particle production. Appealing to the monojet-like signal in the case of such 
invisible decays might be useful to claim a signal excess but would certainly not verify the the origin as due to the production and decay of neutral VL isodoublet PM states of relevance here. 
Unfortunately, the present scenario can provide multiple sources for events with MET so extracting and isolating the signal arising for any particular final state would certainly be quite non-trivial.

\begin{figure}[htbp]
\centerline{\includegraphics[width=5.0in,angle=0]{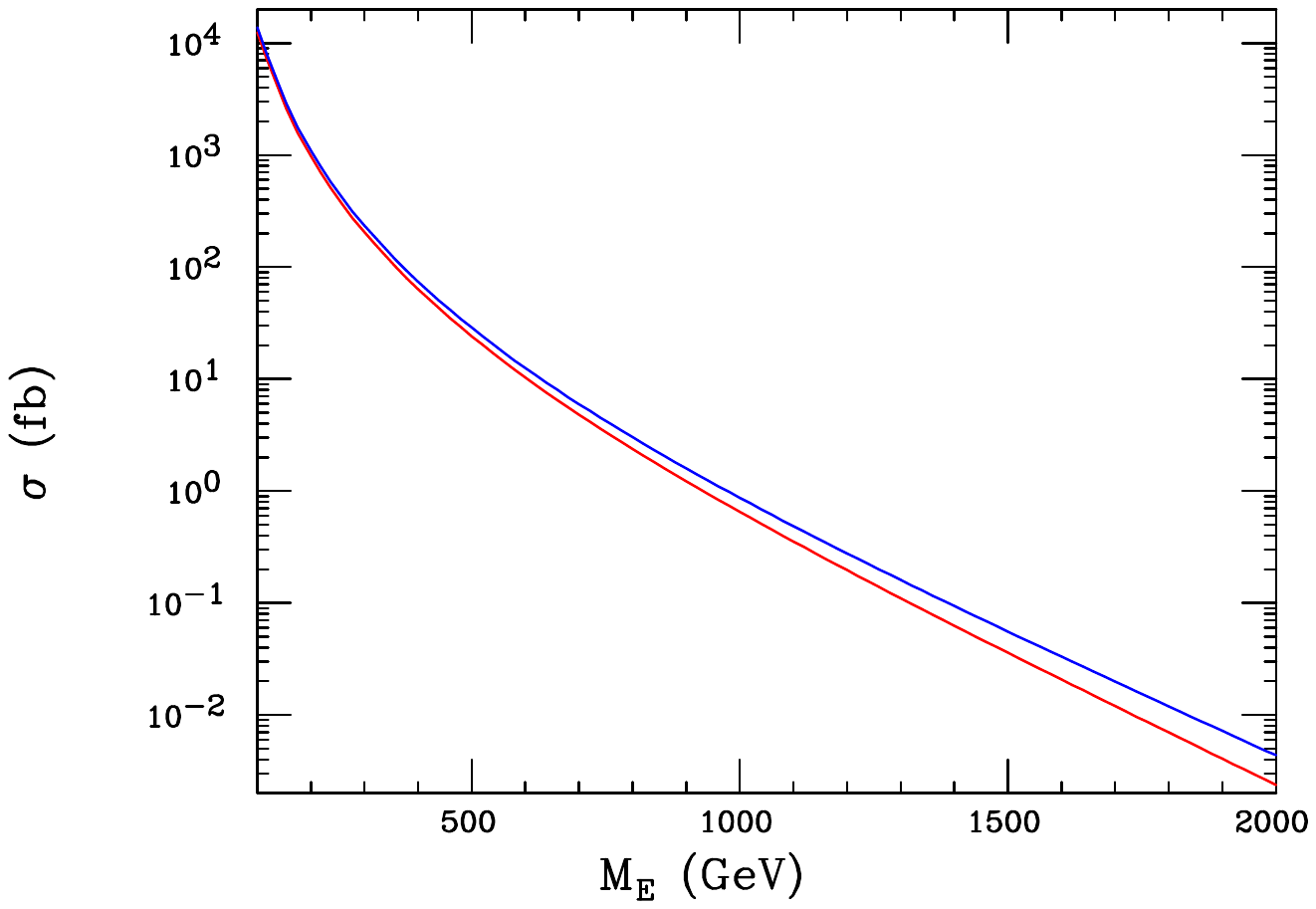}}
\vspace*{-1.2cm}
\centerline{\includegraphics[width=5.0in,angle=0]{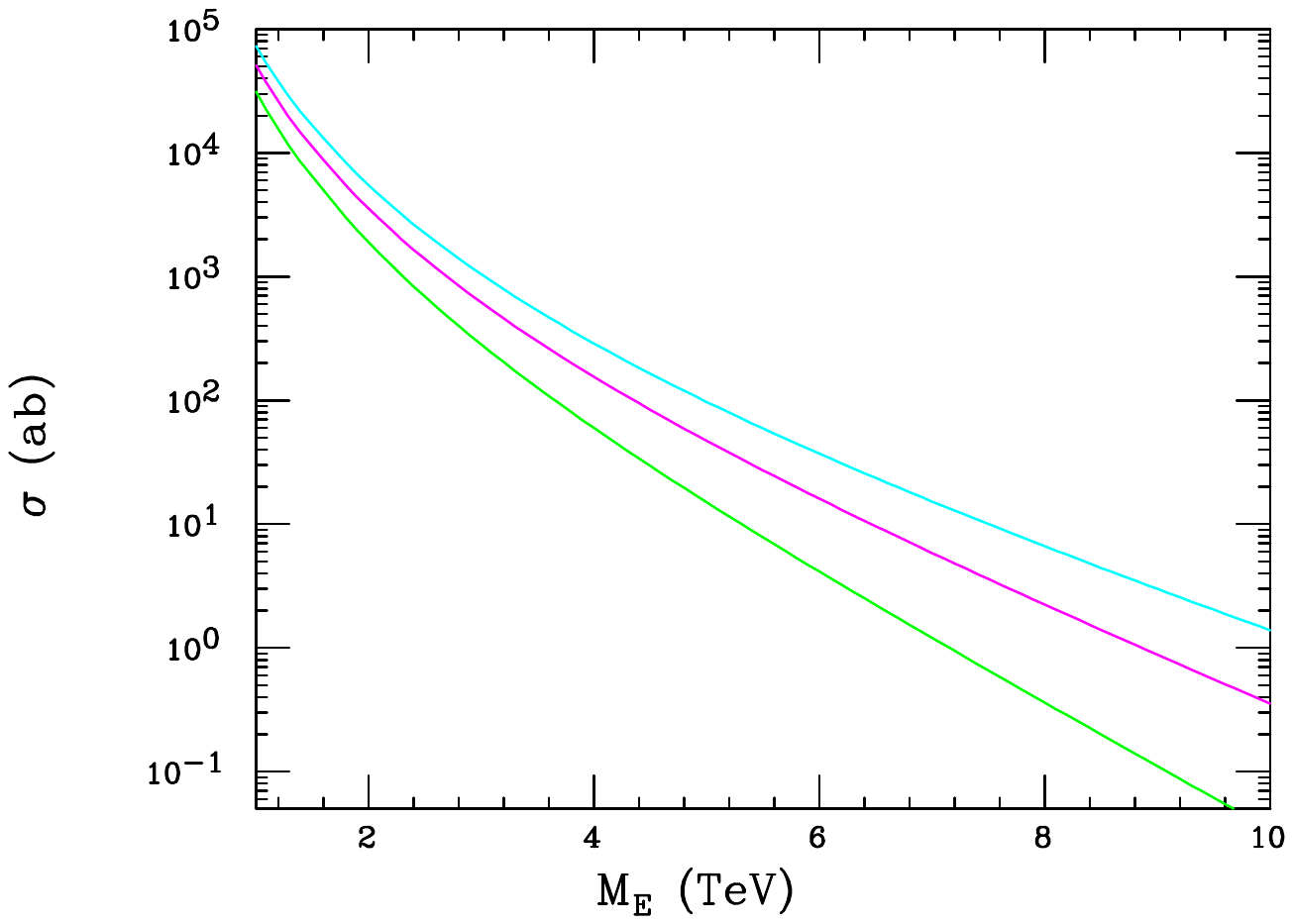}}
\vspace*{-1.3cm}
\caption{Cross sections for associated $\bar NE+h.c.$ production via $s$-channel $W^\pm$ exchange at the LHC (top) with $\sqrt s=13(14)$ TeV and at FCC-hh (bottom) assuming 
$\sqrt s=60, 80,100$ TeV as in the previous Figures.}
\label{fig3}
\end{figure}

In the discussion above we have essentially focussed on a single generation of additional new PM fields or, at least, three simple flavor diagonal copies of a single generation thus ignoring 
any potential flavor issues/constraints. Of course, in the real world, there will be intergenerational mixings (separately) between the full set of $Q=-1$ and $Q=0$ fields as the Yukawa couplings 
effectively gain flavor indices. Thus, in addition to the direct production probes of these new heavy states at colliders, their existence may be probed indirectly via well-known low energy processes, 
\eg, leptonic flavor-changing reactions such as $\mu \to e\gamma$ and $\mu N \to eN$ which are currently under experimental scrutiny.  However, in order to evaluate or at least estimate the 
importance of such effects on constraining the present model parameter space, some realistic flavor symmetry scheme is required to be imposed as the number of potentially contributing states 
is somewhat large which is clearly beyond the scope of the present 
analysis. In the case of {\it flavor diagonal} processes, such as the $g-2$ of the muon, at least in the present setup discussed above, all of the new contributions are helicity suppressed due to the chiral 
couplings of the, \eg, heavy $W_I$ gauge boson and so are suppressed with respect to the SM electroweak contributions by powers of $\sim m_W^2/m_{W_I}^2\lsim 10^{-4}-10^{-3}$ and thus 
essentially invisible.

\section{Discussion and Conclusion}

The new heavy PM fields (together with their dark sector partners) which were introduced to generate the KM between the photon and the DP at low energies can lead to other new physics beyond 
this single interaction, depending upon their specific transformation properties under both $G_{SM}$ and $G_D$. For example, it was shown in earlier work\cite{Rizzo:2021lob}, that 
loops of such fields can generate 
additional new interactions between the various SM fields (for both the fermions and the $W$) with the DP which take the form of anomalous `dark' moments thus leading to both new experimental 
signatures and additional paths by which DM can reach its observed relic density. In this paper, we demonstrated, for the case of a previously introduced $E_6-$like toy model example, how PM 
models can also be employed to explain the absence of tree-level neutrino masses while simultaneously generating a small Dirac neutrino mass at the 1-loop level - employing 
color singlet fermion PM and dark sector fermion and scalar fields - in a manner somewhat similar to that which happens in scotogenic models.  With $O(1)$ Yukawa and quartic couplings,   
the suppression of this mass down to the required $\sim 0.05$ eV scale essentially occurs due the required small magnitudes, $\sim 1$ GeV, of the three $U(1)_D$-breaking Higgs vevs appearing 
in the loop in comparison to the anticipated $\sim$ few TeV masses expected for the color-singlet PM fields, combined with the usual $1/16\pi^2$ loop factor.  Furthermore, from the 
expressions we obtained above, it is clear that such mass values can be realized for a respectable region in the model's parameter space. 

Since most of the fields which participate in the neutrino mass generating loop are neutral color-singlets and/or dark sector fields, it is difficult to directly test this setup in a unique manner, \eg, via 
the production of the relevant new states at colliders, since the particles in question either do not couple to the SM directly and/or they will decay invisibly simply adding into an already existing 
excess of missing $E_T$. An exception to this purely MET signature, but one that will appear in any $E_6-$like model of this kind (which involve lepton-like fields) so that it will not {\it uniquely} test 
the current setup, is the production and decay of the charged SM weak isodoublet PM field, $E$, whose neutral partner, $N$, with which it is (essentially) degenerate, is a necessary participant in 
the loop generating the neutrino Dirac mass.  As was noted, being produced via the $s-$channel exchange of SM electroweak gauge bosons, the HL-LHC has a reach for such fields of the 
first/second generation out to masses of roughly $\simeq 1.6$ TeV employing slepton-like searches. Much larger masses of several TeV may eventually become accessible to the FCC-hh and 
any future multi-TeV lepton collider. This may be a first step in a series of observations that are necessary to test the current scenario. 
One aspect of this specific setup which may be testable at colliders in the future, \ie, that $\nu_R$ carries a dark charge, relies on the examination of the 
production and partial decay widths of the $G_D$ analog of the SM $Z$, $Z_I$. However, we know from current LHC data that the $Z_I$ must be more massive than roughly $\simeq 5$ TeV, 
subject to the assumption that its overall coupling strength is similar to that of the SM $Z$ and that the PM fields are too heavy to appear as final states in its decay. Such measurements are thus 
beyond the capability of the HL-LHC and thus direct collider tests of this general idea will likely require new high energy hadron and/or lepton machines. It is possible and maybe likely that 
astrophysical and cosmological constraints on this setup will prove to be very valuable. 

Hopefully, some signals arising from both the $E_6$-like visible and dark sectors will soon be observed in the laboratory.

\section*{Acknowledgements}
The author would like to particularly thank J.L. Hewett for both hospitality and wide ranging discussions. This work was supported by the Department of Energy, Contract DE-AC02-76SF00515.



\end{document}